\documentclass[11pt]{article}

\usepackage{euscript}
\usepackage{amssymb}
\usepackage{amsfonts}
\usepackage{amsbsy}
\usepackage{amsmath}
\usepackage{epsfig}
\usepackage{amsthm}
\usepackage{amscd}
\usepackage{amstext}
\usepackage{graphicx}

\usepackage[pdftex]{hyperref}

\def\ben{\begin{equation}}
\def\een{\end{equation}}
\def\half{{\textstyle{1\over2}}}
\def\qtr{{\textstyle{1\over4}}}
\let\a=\alpha \let\b=\beta \let\g=\gamma \let\d=\delta 
   \let\k=\kappa
\let\l=\lambda \let\m=\mu \let\n=\nu   
 \let\t=\tau

    \let\L=\Lambda

\def\nn{\nonumber}

\let\pa=\partial
\def\be{\begin{equation}}
\def\ee{\end{equation}}
\def\ba{\begin{array}}
\def\ea{\end{array}}

\def\dalemb#1#2{{\vbox{\hrule height .#2pt
       \hbox{\vrule width.#2pt height#1pt \kern#1pt
               \vrule width.#2pt}
       \hrule height.#2pt}}}

\newcommand{\bea}{\begin{eqnarray}}
\newcommand{\eea}{\end{eqnarray}}

\def\til{\tilde}

\def\CE{{\cal E}}

\def\CF{{\cal F}}
\def\CJ{{\cal J}}
\begin{document}

\begin{center}

\vspace{1cm} { \LARGE {\bf Deformations of
 Lifshitz holography}}

\vspace{1cm}

Miranda C. N. Cheng, Sean A. Hartnoll and Cynthia A. Keeler

\vspace{0.8cm}

{\it Department of Physics, Harvard University,
\\
Cambridge, MA 02138, USA \\}

\vspace{0.6cm}

{\tt  mcheng,hartnoll,cakeeler @physics.harvard.edu} \\

\vspace{2cm}

\end{center}

\begin{abstract}

The simplest gravity duals for quantum critical theories with $z=2$ `Lifshitz' scale invariance admit a marginally relevant deformation.
Generic black holes in the bulk describe the field theory with a dynamically generated momentum scale $\Lambda$ as
well as finite temperature $T$. We describe the thermodynamics of these black holes in the quantum critical regime where
$T \gg \Lambda^2$. The deformation changes the asymptotics of the spacetime mildly and leads to intricate UV sensitivities of the theory which we control perturbatively in $\Lambda^2/T$.

\end{abstract}

\pagebreak
\setcounter{page}{1}

\section{Introduction}

At distance scales large compared to the lattice spacing, condensed matter systems are
often described by quantum field theories (e.g. \cite{Altland}). When the underlying
electronic ground state undergoes a continuous nonanalytic change as a function of an external parameter, such as pressure, doping or magnetic field, the resulting quantum field theory becomes
quantum critical \cite{sachdev}. Quantum critical theories are of interest both because they can
control nonstandard phenomenology over large regions of the phase diagram and also because
they provide an important theoretical starting point that is not necessarily weakly coupled.

The quantum critical theories arising in condensed matter systems are scale invariant, but in general space and time need not scale equally \cite{Hertz1976}. The dynamical critical exponent $z$ determines the relative scaling in which $t \to \lambda^z t$ but $\vec x \to \lambda \vec x$. Given recent excitement about application of the holographic correspondence \cite{Maldacena1998a} to condensed matter systems in general and quantum critical systems in particular \cite{Hartnoll2009, Herzog2009, McGreevy2009, Hartnoll2009b}, it was natural to ask whether the correspondence could be extended to `nonrelativistic', i.e. $z \neq 1$, scale invariance.

Following the logic of the original holographic correspondence, the authors of \cite{Kachru2008}
wrote down a spacetime metric in which the scaling symmetry was realised geometrically:
\be\label{eq:generalz}
ds^2 \sim \ell^2 \left(- \frac{dt^2}{r^{2z}} + \frac{dr^2}{r^2} + \frac{dx^2 + dy^2}{r^2} \right) \,.
\ee
We will focus in this paper on a 3+1 dimensional bulk, with a 2+1 dimensional (putative) dual
field theory. For $z=1$ this spacetime becomes Anti-de Sitter space and the usual AdS/CFT
correspondence is recovered. The full symmetry algebra of the spacetime (\ref{eq:generalz})
is often called the Lifshitz algebra \cite{Adams2009}.
Retarded Green's functions of operators dual to
scalar fields in the bulk metric (\ref{eq:generalz}) were computed in \cite{Kachru2008}
and found to have the form expected for field theories with Lifshitz symmetry.

In order to obtain geometries like (\ref{eq:generalz}),
with anisotropic scaling of space and time, it is clear
that an anisotropic energy-momentum tensor is needed to source the
gravitational field. The minimal way to achieve this is to include a vector field.
It was shown in \cite{Taylor2008} that a massive vector, i.e. Proca \cite{proca},
field is necessary. The Proca field is hodge dual to the two and three form action originally
used in \cite{Kachru2008}. In this paper we will use an Einstein-Proca action.

Given that additional structure beyond pure Einstein gravity is necessary to
have a Lifshitz metric as the ground state, it is important to understand the
physics that is necessarily included in the theory with the addition of a Proca field.
In the holographic correspondence, bulk fields correspond to operators in the dual
field theory. The inclusion of a bulk Proca field indicates that Lifshitz field theories
with gravitational duals come canonically equipped with a preferred vector operator,
call it $J^a$ (not a conserved current). This is in addition to the energy-momentum tensor $T^{ab}$ which is dual to the bulk metric.

The essential physics of the Proca field in the case $z=2$ was explained in \cite{Kachru2008}:
the Proca field (mixing with the graviton) describes a marginally relevant operator in the quantum critical theory. Turning on this operator induces a flow from the $z=2$ theory to a relativistic $z=1$ infrared
fixed point. The main objective of this present paper is to describe the renormalisation group
flow more precisely and in addition explore the physics of the marginally relevant
operator at finite temperature. That is, we are studying a canonical deformation away from criticality and looking at the interplay of the deformation scale and temperature scale.

The value $z=2$ is of particular interest both physically and mathematically. Physically, this value arises at `deconfined' quantum critical points separating different Valence Bond Solid (VBS) phases, e.g. \cite{Rokhsar:1988zz,Vishwanath2009,Fradkin2004,Ardonne2004}. The value $z=2$ was also recently shown to arise in a simple string theory construction involving a distribution of e.g. D0 branes \cite{Hartnoll:2009c}. Mathematically, $z=2$ in the Einstein-Proca system is a special value at which the additional operator due to the Proca field becomes marginal. For $z>2$ the operator is relevant while for $z<2$ it is irrelevant \cite{Bertoldi2009, Ross2009}. In the marginal case, which in fact turns out to be marginally relevant, one finds a dynamically generated scale and nontrivial logarithmic running of various quantities. The renormalisation flow of the bulk fields has similarities to that encountered in `improved holographic QCD' \cite{Gursoy2007} and in topologically massive gravity \cite{Skenderis:2009}. The holographic renormalisation of theories with marginally relevant operators is significantly more intricate than the relevant case that is usually studied; the asymptopia is mildly deformed away from its Lifshitz behaviour, leading to some interesting technical challenges which we partially resolve.

In sections 2 to 4 below we introduce the Einstein-Proca theory and the space of solutions we will be considering. We characterise the asymptotic behaviour of these solutions and show how to compute finite thermodynamic quantities using holographic renormalisation. In section 5 we obtain numerical (planar) black hole solutions and use our previously developed formalism to extract their thermodynamics. Our numerical results are summarised in graphs showing the dependence of thermodynamic variables on the ratio of the dynamically generated scale $\Lambda$ over the temperature.

\section{Einstein-Proca theory and Lifshitz spacetimes}

We will use the following action for a massive vector field coupled to gravity
\be\label{eq:bulkaction}
S = \int d^4x \sqrt{-g} \left( \frac{1}{2 \kappa^2} \left[ R + \frac{10}{\ell^2}\right] - \frac{1}{g^2} \left[\frac{1}{4} F^2 + \frac{2}{\ell^2} A^2 \right] \right) \,,
\ee
with $F=dA$.
The equations of motion following from this action are
\be\label{eq:Ein}
\frac{1}{2 \kappa^2} \left(R_{\mu \nu} - \half g_{\mu\nu} R - {\textstyle \frac{3}{L^2} } g_{\mu\nu}\right)
= \frac{1}{2 g^2} \left( F_{\mu \rho} F_{\nu}{}^\rho - \qtr g_{\mu \nu} F^2\right)
+ \frac{2}{ g^2 \ell^2} \left( A_\mu A_\nu - \half g_{\mu\nu} A^2 \right) \,,
\ee
and
\be\label{eq:Proca}
\nabla_\mu F^{\mu \nu} - \frac{4}{\ell^2} A^\nu = 0 \,.
\ee
The mass term has been tuned (relative to the cosmological constant term)
in order for the Lifshitz metric with $z=2$ to be a solution.
With the normalisations chosen, the Lifshitz metric is
\be\label{eq:Lif1}
ds^2_\text{Lif} = \ell^2 \left(- \frac{dt^2}{r^4} + \frac{dr^2}{r^2} + \frac{dx^2 + dy^2}{r^2} \right) \,,
\ee
while the vector potential is
\be\label{eq:Lif2}
A = \frac{g}{\sqrt{2}}\frac{\ell}{\kappa} \frac{dt}{r^2} \,.
\ee
Both the metric and vector potential are invariant under the `Lifshitz' scaling $\{t,x,y,r\} \to
\{\tilde\lambda^2 t, \tilde\lambda x, \tilde \lambda y, \tilde\lambda r\}$.
With vanishing vector potential we also have the $AdS_4$ solution
\be
ds^2_\text{AdS} = \frac{3}{5}\ell^2 \left(- \frac{dt^2}{r^2} + \frac{dr^2}{r^2} + \frac{dx^2 + dy^2}{r^2} \right) \,.
\ee
It is important to note that there is no gauge invariance of the action (\ref{eq:bulkaction}). Therefore, the dual Lifshitz field theory does not automatically include a global symmetry, and the Lifshitz background should not be thought of as `charged'. Rather, one can think of the divergence of the Proca field (\ref{eq:Lif2}) near the `boundary' $r=0$ as setting the preferred frame of the dual field theory.

In this paper we will be interested in renormalisation group flows and in finite temperature solutions of the Einstein-Proca theory. These involve breaking the scaling symmetry at low energies and therefore we can introduce explicit functions of $r$ in the metric. Thus we take as our Ansatz
\bea\label{eq:Ansatz}
ds^2 & = & \ell^2 \left(- f(r) dt^2 +  \frac{dr^2}{r^2} + p(r) (dx^2 + dy^2) \right) \, , \\
A & = & \frac{\ell}{\k} \,g\, h(r) dt \,.
\eea
With this parametrisation, the Lifshitz solution has
\be
\text{Lifshitz:} \quad f = \frac{1}{ r^4} \,, \quad p = \frac{1}{ r^2} \,, \quad h = \frac{1}{\sqrt{2}} \frac{1}{r^2} \,,
\ee
while the $AdS_4$ solution has
\be
\text{AdS:} \quad f = p = \frac{3}{5}\, r^{-2\sqrt{\frac{3}{5}}} \,, \quad h = 0 \,.
\ee

With the Ansatz (\ref{eq:Ansatz}), the Einstein-Proca equations of motion are equivalent to the following three
nonlinear ODEs for $\{f(r), p(r), h(r)\}$
\bea\nn
-\frac{4 h^2}{f}-\frac{r p'}{p}+\frac{r^2 f' p'}{2 f p}+\frac{r^2 p'^2}{2 p^2}-\frac{r^2 p''}{p}&=&0 \,, \\\label{2nd_order_eqn}
20+\frac{12 h^2}{f}-\frac{r f'}{f}+\frac{r^2 f'^2}{2 f^2}-\frac{2 r^2 f' p'}{f p}-\frac{r^2 p'^2}{2 p^2}-\frac{r^2 f''}{f}&=&0 \,, \\\nn
10+\frac{4 h^2}{f}-\frac{r^2 h'^2}{f}-\frac{r^2 f' p'}{f p}-\frac{r^2 p'^2}{2 p^2}&=& 0 \,.
\eea
Before attempting to solve these equations, it is convenient to make the following change of variables
\be\label{eq:transformation}
p(r) = e^{\int^r q(s) s^{-1} ds} \,, \quad f(r) = e^{\int^r m(s) s^{-1} ds} \,, \quad h(r) = k(r) \sqrt{f(r)} \,.
\ee
This transformation has the virtue of isolating the scaling ambiguity in $p$ and $f$ due to the possibility of rescaling the coordinates $\{t,x,y\}$.
We will postpone fixing this ambiguity until our numerical section 5 below. 
Furthermore, now only first derivatives appear in the equations of motion. A final simplification is achieved by introducing
$$
x = \sqrt{40 + 16 k^2 - 4 m q - 2 q^2} \;,
$$
which leads to the equations of motion
\bea
r x' & = & - 8 k - q x \,, \label{eq:x} \\
r q' & = & 5 - 2 k^2 - \frac{3 q^2}{4} - \frac{x^2}{8} \,, \label{eq:q} \\
r k' & = & - \frac{5 k}{q} - \frac{2 k^3}{q} + \frac{kq}{4} - \frac{x}{2} + \frac{k x^2}{8 q} \label{eq:k} \,.
\eea
In deriving the above equation we have made an assumption about a sign which states that $\frac{d}{dr}A_t<0$, or equivalently the Proca field grows as the boundary is approached. Changing the choice of this sign maps solutions of the Einstein-Proca theory onto solutions with the opposite sign of $A_t$. However our requirement that the sign of the derivative remains the same for all $r$ is significant. A priori the sign could potentially change at some radius on a given solution. We will find that this does not occur on the solutions of interest.

The equations (\ref{eq:x}) to (\ref{eq:k}) are, unsurprisingly, similar to those given in \cite{Kachru2008} for the Hodge dual theory. These are the equations we shall work with.
In terms of these variables, the Lifshitz solution is now simply
\be\label{eq:Lifsimple}
\text{Lifshitz:} \quad q = -2 \,, \quad x = 2\sqrt{2} \,, \quad k = \frac{1}{\sqrt{2}} \,,
\ee
while AdS has
\be
\text{AdS:}
\quad q = -\sqrt{\frac{20}{3}} \;,\quad x=0\;, \quad k = 0  \,.
\ee

From the above equations we can construct an RG-invariant quantity. The combination
\be\label{niftycomb}
K =-\frac{1}{2} \sqrt{f} p \Big(-q+m+k \,x\Big) = -\frac{ \sqrt{f} p}{8 \,q}\Big( 40+16 k^2 -6q^2 +4 q x-x^2 \Big) \;,
\ee
is easily seen to be r-independent. This will turn out to be extremely useful in relating various physical quantities in our solutions.

\section{Asymptotic behaviour and the marginally relevant mode}

To understand the dual field theory physics of the various bulk fields, one starts by expanding near the `boundary' of the spacetime at $r=0$. The form of a general solution to the three first-order equations (\ref{eq:x}) - (\ref{eq:k}) as $r \to 0$ contains three constants $\{\Lambda, \a, \b\}$. One finds
\bea\notag
k& =&\frac{1}{\sqrt{2}}\left(1+ \frac{1}{\log(\L r)} +\frac{(-3+\l)-5\log(-\log(\L r))}{2\log^2(\L r)} +\dotsi\right) \\ \notag
&+& (\L r)^4  \log^2(\L r)  \left(\b\left( 1+\frac{5 \log(-\log(\L r) )}{\log(\L r) }+\dotsi\right) +{\a}\left(\frac{1}{\log(\L r)}+ \dotsi\right)\right)  +{\cal O}(r^8) \,, \\\label{series_solutions}
q&=& -2\left( 1 - \frac{1}{\log(\L r)} -\frac{(1+\l)-5\log(-\log(\L r))}{2\log^2(\L r)} +\dotsi\right) \\\notag
&-&\frac{2\sqrt{2}}{3} (\L r)^4  \log^2(\L r) \left(\b\left(1+\frac{-\frac{4}{3}+5{\log}(-\log(\L r) )}{\log(\L r) }+\dotsi\right) +{\a}\left(\frac{1}{\log(\L r)}+ \dotsi\right)\right)  +{\cal O}(r^8) \,, \\ \notag
x&=&2\sqrt{2}\left(1+ \frac{2}{\log(\L r)} +\frac{\l-5\log(-\log(\L r))}{\log^2(\L r)} +\dotsi\right)\\\notag
&-&\frac{8}{3} (\L r)^4  \log^2(\L r) \left(\b\left(1+\frac{-\frac{7}{3}+5{\log}(-\log(\L r))}{\log(\L r) }+\dotsi\right) +{\a}\left(\frac{1}{\log(\L r)}+ \dotsi\right)\right)  +{\cal O}(r^8)\;,
\eea
where the ellipses denote the terms suppressed by a factor of $\frac{\log(-\log(\L r))}{\log(\L r)}$ or more. Notice that the above are UV expansions in that they are valid near the boundary when $\L r \ll 1$. We have not linearised the equations.

Some comments about the parameters in the above expressions are in order.
It appears that, apart from the three parameters $\{\Lambda, \a, \b\}$ controlling the three modes, there is an extra parameter $\l$ that enters the solution. In fact the presence of $\l$ does not signal another degree of freedom in our system. Rather it is related to a `gauge choice' in defining the scale $\L $. More precisely, the solutions are mapped on to each other by the following transformation
\be\label{covariance1}
k(\L r;\a,\b;\l) = k(e^{\l'/2} \L r; e^{-2\l'}(\a-\l' \b),e^{-2\l'} \b; \l+\l')\;,
\ee
with the same holding for the other two functions $q,x$.
This fact can be seen by verifying that all the series in the above solutions (\ref{series_solutions}) can be obtained by putting $\l$ to zero and at the same time replacing $\log(\L r)$ with  $\log(e^{-\l/2}\L r)$ and then re-expanding the series under the assumption $|\log(\L r)| \gg |\l|$.
In order to compare different solutions, we can fix this ambiguity by fixing the parameter $\l$ once and for all. Subsequently, the parameters $\{\L,\a,\b\}$ have an unambiguous meaning and can be compared among different solutions. From now on we will take $\l=0$.

We see that there is one mode with an inverse logarithmic falloff and then two modes that go to zero as $r^4$ times logarithms. If we count the overall scale for the metric mode $g_{tt}$, left unfixed in (\ref{eq:transformation}), then up to logarithms we can think of this asymptotic behaviour as two `non-normalisable' modes that tend to a constant together with two `normalisable' modes that go to zero. In the holographic correspondence, each operator in the field theory will be dual to a pair of modes. The `non-normalisable' mode specifies the source for the operator while the `normalisable' mode determines the vacuum expectation value. The power of the normalisable falloff determines the scaling dimension $\Delta$ of the dual operator.

The scaling dimension $\Delta=4$ implies that we have two marginal operators in $z+2=4$ effective spacetime dimensions of the field theory. One of these will be the energy density $T^{tt}$, dual to $g_{tt}$, which necessarily has mass dimension $[T^{tt}] = z+2=4$. This mode is marginal for all $z$. The other operator $J^t$, however, is only marginal for $z=2$ \cite{Bertoldi2009, Ross2009}. The expansions above (\ref{series_solutions}) indicate that there is a logarithmic mixing between the two marginal operators.

The inverse logarithmic falloff in (\ref{series_solutions}) is consistent with a marginally relevant operator, whose renormalisation group flow dynamically generates the energy scale $\L^2$ (cf. \cite{Gursoy2007}). Recall that $\Lambda$ is a momentum, so $\Lambda^2$ is an energy in a theory with $z=2$. The fact that the operator is marginally relevant means that the Lifshitz scaling will be approximately recovered in the far UV, at energies much greater than the dynamical energy scale $\Lambda^2$. We will quantify this statement very shortly as it turns out to be a little subtle. Previous works on black holes in the Proca-Einstein theory have tuned the marginally relevant mode to zero, e.g. \cite{Taylor2008, Bertoldi2009, Ross2009, Mann2009, Danielsson2008, Bertoldi2009a}. This is the choice to work precisely at the quantum critical point. At the critical point the holographic renormalisation of the theory is significantly easier, as the slow falloff of the inverse logarithm term introduces many new volume divergences of the on-shell action.

In this paper we wish to consider the more generic solution in which the marginally relevant coupling is allowed to run. This will ultimately allow the study of crossover scaling physics as a function of $T/\Lambda^2$ (cf. \cite{sachdev}). The complication is that it will be challenging to holographically renormalise the logarithmic running. We shall approach this problem by working perturbatively in an expansion in inverse powers of $ \log \frac{T}{\Lambda^2}$. Concretely, the temperature will be much larger than the dynamically generated scale.

Before turning to the issue of holographic renormalisation, we should look more closely at the asymptopia of the spacetime.
From the expansions (\ref{series_solutions}) and (\ref{eq:transformation}) we get
\bea \notag
-\frac{1}{\ell^2}g_{tt} (\rho)&=& F_0^2 \, \frac{1}{(\L r)^4 \log^4(\L r)}
\left( 1+\frac{-10-10{\log}(-\log(\L r))}{\log(\L r) }
+\dotsi\right) \,, \\\label{asym_metric}
\frac{1}{\ell^2}g_{xx} (\rho)&=& P_0^2 \, \frac{\log^2 (\L r)}{(\L r)^2} \left(1+\frac{4+5{\log}(-\log(\L r) )}{\log(\L r) }+\dotsi\right) \;.
\eea
where $F_0, P_0$ are constants. In this expression we see that in our Fefferman-Graham-like coordinates the asymptotic metric has logarithmic deformations away from Lifshitz behaviour at leading order \cite{Danielsson2008}. The curvature invariants and spacetime volume element $\sqrt{-g}$ in contrast are found to tend to their Lifshitz values plus inverse logarithmic corrections. This raises the question of whether these solutions can be included in an `asymptotically Lifshitz' holographic correspondence. This concept has not been fully defined yet (some initial developments appeared in \cite{horava2009}) and will be an important question for future work. Analogous, though weaker, slowly falling off modes due to marginal deformations in the better characterised asymptotically AdS setup can be found for instance in \cite{Henneaux2004, Amsel2006} . In the following we will present various arguments suggesting that one can make sense of the asymptopia (\ref{asym_metric}), such as the computation of finite expectation values for operators including the energy. Our underlying reason for believing this behaviour should be allowed is that the UV deformation (\ref{asym_metric}) was found in \cite{Kachru2008} to lead to a renormalisation group flow into the IR with the properties one should expect of a flow generated by a (marginally) relevant operator. We will find below that the finite temperature physics is also strongly consistent with this picture; for instance the deformation becomes increasingly less important at high temperatures. Marginally relevant deformations do not render the UV theory ill-defined. The logarithmic scaling in (\ref{asym_metric}) would therefore be a manifestation of the fact that at any arbitrarily large fixed energy there is still a logarithmic deviation from scale invariance, as occurs for instance in QCD.

To disentangle various effects, and simultaneously connect with previous results at $\Lambda = 0$, suppose we are interested in the behaviour of the series expansions at some high energy scale $\frac{1}{r} \gg \L$. Following the usual renormalisation group logic we can introduce a sliding scale $\mu \gg \L$ such that $\mu r \lesssim 1$. Then, we have
\be\label{eq:split}
\log \L r = - \log \frac{\mu}{\L} + \log \mu r \,.
\ee
By expanding in
\be
\left|\frac{1}{\log \frac{\mu}{\L}}\right|\,, \;\left|\frac{\log \mu r}{\log \frac{\mu}{\L}}\right| \ll 1 \,,
\ee
 we can read off the leading order running of the couplings of the theory under the renormalisation group.
What we are doing here is formally showing how to take the $\Lambda \to 0$ limit, which does not commute with the $r \to 0$ limit.
We first expand in $\frac{1}{\log \frac{\mu}{\L}}$ keeping $\mu r$ fixed. We can then take $\mu r$ small if we wish.
When we later turn to high temperatures, $T \gg \L^2$, we will find that the re-expansion with $T \sim \mu^2$ is accurate over a significant
range of $r$. In order for this re-expansion to make sense, we need to keep the right quantities fixed as we turn off the dynamical scale $\Lambda$. Specifically we let
\bea\notag
\til \a &=& \frac{\Lambda^4}{\mu^4} \left[ -{\log}\left(\frac{\mu }{\Lambda }\right)\,\a+{\log}\left(\frac{\mu }{\Lambda }\right)\,\left({\log}\left(\frac{\mu }{\Lambda }\right)-\frac{5}{2}\log {\log}\left(\frac{\mu }{\Lambda }\right)+ \frac{1}{12}\right)
\,\beta \right] \,, \\
\til \b&=& \frac{\Lambda^4}{\mu^4} \left[ \alpha -2\,{\log}\left(\frac{\mu }{\Lambda }\right)\,\beta \right]   \;.
\eea
This redefinition allows us to take consistently  the $\mu/\L \to \infty$ limit of the expansions (\ref{series_solutions}) with $\til\a$, $\til\b$ and $\mu r$ held fixed.

This re-expansion, at the leading order, leads to exactly the linearized approximation to the differential equations  (\ref{eq:x})-(\ref{eq:k})
about the Lifshitz backround in the absense of the marginally relevant deformation,
as was employed in earlier works such as \cite{Kachru2008, Ross2009, Bertoldi2009}.
Concretely, re-expanding our general solutions (\ref{series_solutions}) in the way described above in terms of the new parameters leads to
\bea\notag
k &=& \frac{1}{\sqrt{2}} + (\m r)^4 \big(  \til\b \log(\m r)+\til\a\big)+\dotsi+ {\cal O}((\m r)^8) \,, \\
\label{double_sol}
q&=& -2 - \frac{2\sqrt{2}}{3}\, (\m r)^4\, \big(  \til\b \log(\m r)+\til\a-\til\b\big)+\dotsi + {\cal O}((\m r)^8) \,, \\\notag
x&=& 2\sqrt{2}-\frac{8}{3}\, (\m r)^4\,\big(  \til\b \log(\m r)+\til\a\big)+\dotsi+ {\cal O}((\m r)^8)\;,
\eea
where the ellipsis denotes terms that are suppressed by factors of $\frac{\log \log(\m/\L) }{\log(\m/\L)}$  or $\frac{1}{\log(\m/\L)}$.
The higher order terms can be found systematically.

We can now return to the asymptotics of the geometry.  After re-expanding the metric (\ref{asym_metric}) in the above way, we can also rescale  the $t,x$ coordinates as
\be\label{eq:rescaling}
t \to \frac{(\L \log(\frac{\m}{\L}))^2}{F_0} \,t\;,\quad x\to\frac{\L}{\log(\frac{\m}{\L})}\frac{1}{P_0} \,x\;,
\ee
after which the metric takes the form
\bea\label{asymptotics}
-\frac{1}{\ell^2}\,g_{tt} (\rho)&=& \frac{1}{ r^4} \Big(1+4\frac{\log(\m r)}{\log(\frac{\m}{\L})}+\dotsi\Big) \,, \\ \label{asymptotics2}
\frac{1}{\ell^2}\,g_{xx} (\rho)&=&\frac{1}{ r^2} \Big(1-2\frac{\log(\m r)}{\log(\frac{\m}{\L})}+\dotsi\Big)\, .
\eea
The Proca field behaves similarly, allowing us to
conclude that in the regime where the re-expansion is valid the spacetime can be treated as asymptotically Lifshitz with parametrically small deviations. Because $\mu$ is arbitrary, we can perform the re-expansion at arbitrarily high energies, consistent with the notion that at any fixed high energy the spacetime is Lifshitz plus a small breaking of scaling symmetry due to a marginally relevant operator. When we heat up the system, it will be natural to consider this argument with $\mu^2 \sim T$. In the limit $T/\Lambda^2 \to \infty$, precise Lifshitz invariance is recovered at energy scales much greater than the temperature.
The various scales of interest are shown in figure \ref{scaleplot} below.

\begin{figure}[h]
  \begin{center}
    \includegraphics[width=5in]{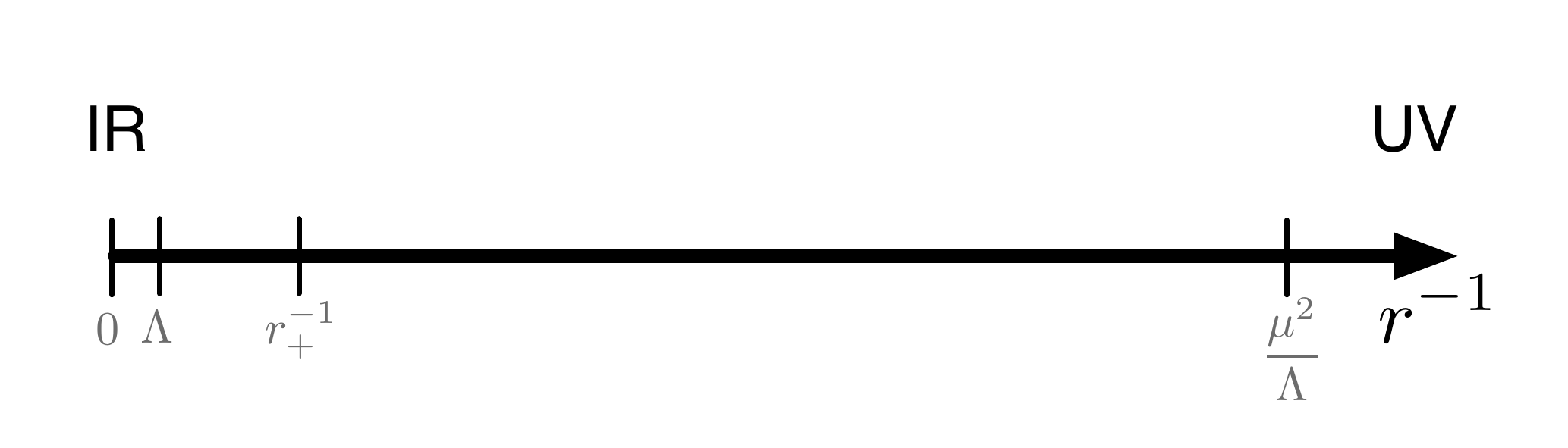}
  \end{center}
   \caption{\small A few scales of significance in our discussion and solutions. The first one is $\Lambda$, which is the scale dynamically generated by the marginally relevant mode. In particular $\Lambda=0$ when the marginally relevant mode is turned off. For the high temperature black hole solutions we are able to discuss most explicitly, the scale $\Lambda$ is behind the event horizon located at $r_+^{-1}$. To the right is the sliding UV cut-off beyond which the strict Lifshitz asymptopia is destroyed (for a given choice of $\mu$, which can be taken arbitrarily large). }
  \label{scaleplot}
  \end{figure}

One reason that the re-expansion is technically useful is that it moves the logarithms of $r$ from the denominator in (\ref{series_solutions}) to the numerator in (\ref{double_sol}). This is again familiar from QCD. To any fixed order in perturbation theory, logarithms of the scale in the renormalisation flow of the coupling will only appear in the numerator. However, because we know that the QCD coupling is asymptotically free, these logarithms must be resummed into the denominator so that the coupling flows to zero in the UV. Our original expansions (\ref{series_solutions}) are the resummed expressions, while (\ref{double_sol}) has reintroduced a perturbation theory in the marginally relevant coupling by working at a scale $\mu$ which is much higher than the dynamically generated scale $\Lambda$. Similarly, at temperatures $T \gg \Lambda^2$ we will find that the deformation can be treated perturbatively.

Before moving on to discuss holographic renormalisation we briefly recall the physical role of relevant deformations in a weakly coupled realisation of $z=2$ Lifshitz criticality and a difference with the present strongly coupled case. The critical Lifshitz action considered in \cite{Vishwanath2009,Fradkin2004,Ardonne2004} is, up to `instanton' terms and a marginally irrelevant coupling,
\be\label{eq:weakly}
S_{\text{Lif}} = \half \int d\tau d^2x \left( (\pa_\tau \phi)^2  + \tilde K (\nabla^2 \phi )^2  \right) \,.
\ee
This theory clearly has a $z=2$ scale invariance. The above action admits a relevant deformation by the term $\rho (\nabla \phi)^2$. If $\rho$ is positive then the theory flows to a Lorentz invariant theory in the IR. This is very analogous to the renormalisation group flow found numerically in \cite{Kachru2008} and that we are investigating in more depth in this paper. The difference between the two theories is that in the strongly coupled theory with a holographic dual, the natural deformation is marginally relevant rather than simply relevant as in the weakly coupled case. If $\rho$ is negative then one obtains a `tilted' phase for the `height field' $\phi$, i.e. an expectation value for $\nabla \phi$, that spontaneously breaks spatial isotropy. The Lifshitz theory (\ref{eq:weakly}) is then understood as the quantum critical theory separating tilted and untilted phases at zero temperature. In this paper we are only investigating the approach to criticality from one phase. We hope our results will provide a framework for a holographic realisation of the full quantum phase transition.

While we expect our bulk theory to be dual to a field theory with the same scaling symmetry as the above theory (\ref{eq:weakly}), we do not expect it to be dual to a theory with such a simple Lagrangian with a single scalar field. In the present paper we will not attempt to construct the dual field theory. Rather it will suffice to point out that there is a natural family of theories that has the counter-part of a matrix large $N$ limit. Generalisations of this theory are candidates for the field theory dual to the bulk theory discussed in the present paper.  To motivate the existence of such a large $N$ theory, it will be illuminating to  rewrite the Lagrangian  (\ref{eq:weakly}) in the following way. Recall that a scalar is dual to a one-form in (2+1)-dimensions.
Writing
$$
\star d \phi = d {\bf A}\;,
$$
the above Lagrangian is transformed into
$$
 S_{\text{Lif}} = \half \int d\tau d^2x \left( |{\bf B}|^2  + \tilde K |\nabla \times {\bf E} |^2  \right) \,.
$$
The $SU(N)$, $SO(N)$, $Sp(N)$ generalisations of the above $U(1)$ Lagrangian have been constructed in \cite{Horava2008} and it would be interesting to establish the possible duality between such theories and the various bulk theories, e.g. \cite{Kachru2008,Balasubramanian2009}, which produce the Lifshitz spacetime (\ref{eq:Lif1}) as a solution.

\section{Holographic renormalisation}

Using the equations of motion (\ref{2nd_order_eqn}) it is possible to show that the bulk action (\ref{eq:bulkaction}) evaluated on-shell can be written as the integral of a total derivative and hence becomes a boundary term. Specifically, the Euclidean on-shell action gives the semiclassical free energy density
\be\label{actiondensity}
{\cal F}_E = - \frac{1}{2}  \left(\frac{\ell}{ \kappa}\right)^2 \lim_{r \to 0} \sqrt{f} r p' \,.
\ee
By density we mean the density per unit boundary volume. That is, the free energy is $F = \int d^2x \, {\cal F}$, with no integral over the Euclidean time circle.

Before evaluating this action, we will need to add the boundary action and counterterms. Firstly there is the usual Gibbons-Hawking term required for the desired variational principle, which in Euclidean space evaluates to
\be
{\cal F}_\text{G-H} = \frac{1}{\k^2} \lim_{r \to 0}   \sqrt{\gamma}\,K
=  \left(\frac{\ell}{ \kappa}\right)^2 \lim_{r \to 0} r (\sqrt{f} p)' \,,
\ee
where $\gamma_{ab}$ is the induced metric on the boundary and $n^\m \partial_\m$ is an outward-pointing unit vector to the boundary which defines the extrinsic curvature $K_{\m\n} = -\nabla_{(\m}n_{\n)}$. The bulk and Gibbons-Hawking terms do not lead to a finite on-shell action, and therefore boundary counterterms are required to render the theory well defined.

There has been some confusion in the literature over the correct holographic renormalisation of asymptotically Lifshitz solutions to Einstein-Proca theory. The physics seems clear: the solutions describe two deformations of the Lifshitz theory by the operators $T^{tt}$ and $J^t$ and their corresponding expectation values. As neither of these operators appear to be irrelevant, they do not destroy the UV definition of the theory. Therefore we should be able to consistently describe the physics of all of these modes. It becomes a question of finding the correct boundary counterterms.

The approach we take is similar to \cite{Balasubramanian2009}. In some regards our situation will be easier than the situation in that paper, because the theory studied there leads to unusually slowly falling off modes\footnote{Specifically, \cite{Balasubramanian2009} have metric modes falling off like $r^2$ as opposed to $r^4$ in the coordinates (\ref{eq:Lif1}).} even without the marginally relevant operator. Just as in that work, in order to preserve boundary covariance, we would like to consider
counterterms in the form of a power series in $A^2 = A^\m A_\m$. See \cite{Ross2009} for an alternative choice of counterterms involving the combination $\sqrt{ A^\m A_\m}$, which does not seem to generalise straightforwardly to include the marginally relevant deformation. As we will explain in more detail shortly, it turns out that just three such terms, $\sqrt{\g} (\tilde c_0+ \tilde c_1 A^2  + \tilde c_2 A^4)$, are enough to render all the physical quantities of interest to us finite. The complication due to the marginally relevant deformation is that the coefficients $\{\tilde c_i\}$ become functions of $\log \Lambda r$.

Consider the following boundary counterterms in the free energy density
\bea\notag
{\cal F}_\text{c.t.}  & = &\frac{1}{2\ell \k^2} \lim_{r \to 0}  \sqrt{\gamma} \sum_{n=0}^2 c_n \left(- \frac{\k^2}{g^2} A^2 - \frac{1}{2} \right)^n \\\label{eq:ct}
& = & \frac{1}{2}   \left(\frac{\ell}{ \kappa}\right)^2  \lim_{r \to 0} \sqrt{f} p \sum_{n=0}^2 c_n (k^2-\half)^n \,.
\eea
We chose  to write the series in this form for convenience, since $- \frac{\k^2}{g^2} A^2 - \frac{1}{2} =0$ when evaluated on the Lifshitz solution.
We will shortly write down the coefficients $\{c_i\}$ that make the total semiclassical free energy density
\be\label{eq:Ftotal}
\CF = {\cal F}_\text{E} + {\cal F}_\text{G-H} + {\cal F}_\text{c.t.} =
\frac{1}{2}\left(\frac{\ell}{ \kappa}\right)^2   \lim_{r \to 0} \sqrt{f} p \left(\frac{r f'}{f}+\frac{r p'}{p} + \sum_{n=0}^2 c_n (k^2-\half)^n\right) \, ,
\ee
finite.

Besides rendering the on-shell action finite, the counterterms should also lead to a well-defined variational problem and in particular finite on-shell expectation values and charges obtained by variation of the action with respect to boundary fields.
Na\"ively, following the usual procedure we would write the on-shell variation as
\be\label{naive}
\d \CF  = \frac{\sqrt{\g}}{2}\tau^{ab}\d\g_{ab}+\CJ^a \d A_a\;,
\ee
but some care is needed in the present case.

As shown in \cite{Hollands2005} in the more established AdS/CFT context, the usual charge defined by $\int d^2 x \sqrt{\sigma} \xi_a k_b\tau^{ab} $, where $\xi^a \pa_a$ is a boundary Killing vector, $\sqrt{\sigma} = \sqrt{\gamma_{xx} \gamma_{yy}}$ is the spatial volume element and $k^a \pa_a$  is the unit normal to the boundary Cauchy surface, say $t = \text{const.}$, is not conserved when non-scalar matter fields are present. Instead, following
\cite{Hollands2005,Ross2009}, when defining the energy-momentum tensor we should hold fixed the matter fields in the boundary vielbein frame, defined by the usual relation
\be
\g_{ab} = \eta_{\hat a \hat b} e_a^{\hat a} e_b^{\hat b}\quad,\quad \eta = \text{diag}(\pm1,1,1)\;,
\ee
where we have used the indices $a,b,...$ to denote the local coordinates on the boundary and the hatted indices to denote the local tangent space coordinates.
The boundary stress tensor ${\cal T}^{ab}$ is then given by
\be
\d \CF  ={\sqrt{\g}}\,{\cal T}^{a}_{\;\;\;\hat a}\,\d e_a^{\hat a}+\CJ^{\hat a} \d A_{\hat a}\, ,\qquad
{\cal T}^{ab} = {\cal T}^{a}_{\;\;\;\hat a} e^{b\hat a}\;.
\ee
A short computation shows that
\be
{\cal T}^{ab} = \tau^{ab}+ \frac{1}{\sqrt{\g}}\,{\cal J}^{(a} A^{b)} \,,
\ee
where the bracket denotes symmetrisation of indices. In particular, the energy density is given by
\be
{\cal E} = \sqrt{\sigma}\,  k_a \,\xi_b\, {\cal T}^{ab}= \sqrt{\gamma}\, \tau^t{}_t+  \CJ^t A_t \,.
\ee
While the $\CJ^t A_t$ term here is closely analogous to the standard chemical potential times charge term
appearing in the canonical ensemble, we should remember that $\CJ^t$ in our system is not a conserved charge.

Taking the local frame to be
\be
e^{\hat t}  = e^{\hat t}_a dx^a =  \sqrt{f} d\t\;,\quad e^{\hat x} = \sqrt{p} \,dx \;,\quad e^{\hat y}=\sqrt{p} \,dy\;,
\ee
the quantities of particular interest can be computed to give
\bea
\t^{a b} & = & \frac{2}{\sqrt{\gamma}} \frac{\delta {\cal F}}{\delta \gamma_{a b}} = \frac{1}{\kappa^2} \left(K^{ab}- K \gamma^{ab} \right)  \\
& + & \frac{1}{2 \ell \k^2} \sum_{n=0}^2 c_n \left(\g^{ab}\left(- \frac{\k^2}{g^2} A^2 - \frac{1}{2} \right)^n + 2 n \frac{\k^2}{g^2} A^a A^b \left(- \frac{\k^2}{g^2} A^2 - \frac{1}{2} \right)^{n-1}  \right) \,,
\eea
and
\bea\notag
{\cal J}^{\hat t} &=& \sqrt{f}\frac{\d {\cal F}}{\d A_t} = \frac{1}{g} \frac{\ell }{\k}  \lim_{r \to 0} \sqrt{f} p \left( \frac{ r(\sqrt{f}k)'}{\sqrt{f}} +k \sum_{n=0}^2  n \,c_n  (k^2 - \half)^{n-1} \right)\\\label{gauge}
&=& \frac{1}{g} \frac{\ell }{\k} \lim_{r \to 0} \sqrt{f} p \left( -\frac{1}{2}x +k \sum_{n=0}^2  n \,c_n  (k^2 - \half)^{n-1} \right)\;,
\eea
and all other $\CJ^{\hat a}=0$. We have used the equation of motion (\ref{eq:k}) to obtain the second line of the previous equation. Putting these expressions together, we obtain
\be\label{energy}
\CE = \left(\frac{\ell }{\k} \right)^2 \lim_{r \to 0} \sqrt{f} p \left(  \frac{r p'}{p}-\frac{x k }{2}
+\frac{1}{2}\sum_{n=0}^2 c_n\, (k^2-\half)^n \right)\;.
\ee
It will be very useful to note that the above expression for the energy density is related to the free energy in a simple way. Observe that
\bea
\sqrt{f} p \left(  \frac{r p'}{p}-\frac{x k }{2}
+\frac{1}{2}\sum_{n=0}^2 c_n\, (k^2-\half)^n \right) = \frac{1}{2}   \sqrt{f} p \left(\frac{r f'}{f}+\frac{r p'}{p} + \sum_{n=0}^2 c_n (k^2-\half)^n\right) + K\;,
\eea
where $K$ is the RG-invariant quantity we found in (\ref{niftycomb}). Evaluating both sides in the UV ($r \to 0$) and using (\ref{energy}) and (\ref{eq:Ftotal}) leads to the simple relationship
\be\label{1st_law_1}
\CE= \CF + \left(\frac{\ell}{\kappa}\right)^2\,K \;.
\ee
As we shall see later, this gives exactly the `integrated first law of thermodynamics' when considering a black hole solution.
Finally we can easily compute the pressure and find ${\cal P} = - \sqrt{\gamma} \tau_{\;\;x}^x = - \CF$ as expected on general thermodynamic grounds.
The off-diagonal components such as ${\cal T}^{t}_{\;\;x}={\t}^{t}_{\;\;x}$ vanish on our solution.

As we have implied in the preceding discussion, in order to renormalise the on-shell action we will need three boundary terms given by three non-vanishing coefficients $c_0,c_1,c_2$. It turns out that instead of being constant, they are series in $\frac{1}{\log(\L r)}$, with $r$ understood to be evaluated at a small cutoff value. This explicit dependence of the boundary counterterms on the cutoff is expected: the marginally relevant perturbation of the theory breaks scaling invariance at any fixed scale.
This breaking of the scaling symmetry can be viewed as a conformal anomaly Ê\cite{Henningson1998}. However, scale invariance should be recovered in the strict $r \to 0$ limit, or equivalently the $\L\to 0$ limit, and hence we can expect inverse powers of $\log \Lambda r$ to appear. Imposing finiteness of the free energy and energy densities we find
\bea\nn
c_0&=&6-\frac{8}{3 \log^2(r \Lambda )}+\frac{65 +120 \log(-\log(r \Lambda ))}{9 \log^3(r \Lambda )}+\dotsi \,, \\\label{eq:coeffs}
c_1&=&2+\frac{16}{3 \log(r \Lambda )}+\frac{-91 -120 \log(-\log(r \Lambda ))}{9 \log^2 (r \Lambda )}+\dotsi \,, \\\nn
c_2&=&-\frac{5}{3}+\frac{35}{9 \log(r \Lambda )}+\frac{a-\frac{175}{18} \log(-\log(r \Lambda ))}{\log^2 (r \Lambda )}\;,
\eea
where the ellipses again denote terms suppressed by extra factors of $\log(-\log \Lambda r)/\log \Lambda r$ or $1/\log \Lambda r$.  A few remarks about these coefficients are in order.
Firstly, notice that the first two coefficients are given in terms of infinite series, while the series expansion of the third coefficient terminates at $1/\log^2 (\Lambda r)$. This can be understood in the following way. Observe that the divergences of the free energy and energy before renormalisation are both given in terms of $r^{-4}$ multiplied by an infinite series in $1/\log \Lambda r$, plus a finite number of divergent terms of the form $\log^n(\L r)$ for some positive integer $n$. Hence we have altogether two infinite series and a few extra terms which need to be cancelled by the counterterms. Unsurprisingly, they are exactly cancelled by counterterms given by two infinite series and three more terms.  Relatedly, there is an ambiguity in the final counterterm given by a real number denoted by $a$ in the above equation. This ambiguity does not afflict any of the thermodynamical quantities we shall study later. Nevertheless it does represent an ambiguity in the quantity $\CJ^{\hat t}$, which reflects the extent to which the system reacts upon changing the boundary (i.e. background) value of the Proca field $A_{\hat t}$. Such ambiguities are physical and common in the presence of logarithmically running couplings.
It is also possible to add more subleading terms in the series for $c_2$, but these terms will not affect our results for $\CF,\CE$ or ${\cal J}^{\hat t}$. Finally, it is also possible to add higher power counterterms such as $c_3 (k^2-\half)^3$. Nevertheless we shall make the above minimal choice which will be sufficient for our purposes.

After incorporating the appropriate counterterms given by (\ref{eq:ct}) and (\ref{eq:coeffs}), we obtain the following finite expressions for the physical quantities in terms of the parameters $\a,\b$ in our expansions (\ref{series_solutions}):
\bea\nn\CF&=& \left(\frac{\ell}{ \kappa} \right)^2 \frac{\sqrt{2}}{9} \left(-5\b + 6\a \right) \,, \\
\CE&=& - \left(\frac{\ell}{ \kappa} \right)^2 \frac{\sqrt{2}}{9} \left(7\b + 6 \a \,\right) \,, \label{eq:niceEF} \\\nn
\CJ^{\hat t}&=& \frac{1}{g} \frac{\ell}{ \kappa} \left(\frac{27}{703}+2 a\right)\b\;.
\eea
From here we see that evaluated on the pure Lifshitz solution we have
$$
{\cal E}={\cal F}=\CJ^{\hat t}=0 \,.
$$
Furthermore, recall that we have an RG-invariant quantity (\ref{niftycomb}) which can be written in terms of the parameters $\a,\b$ as
\be\label{xintermsofalphabeta}
K = -\frac{1}{2} \sqrt{f} p \Big(-q+m+k \,x\Big)  =-\frac{2\sqrt{2}}{9} \left(\b + 6 \a\right)\;,
\ee
consistent with the relation (\ref{1st_law_1}) between the quantities $\CE$, $\CF$ and $K$.

\section{Finite temperature}

Several works have studied black holes in Einstein-Proca theory for various values of $z$, e.g. \cite{Bertoldi2009, Mann2009, Danielsson2008, Bertoldi2009a}. As we noted above, these studies focused on the quantum critical theory and tuned the marginally relevant deformation to zero. We will begin similarly to those studies, by expanding the (planar) black hole solution near the horizon, and then numerically integrating outwards towards the boundary.  Since we have now learned how to analyze solutions which contain the marginally relevant mode at the boundary, we will be able to explore the energetics of these black objects for a family of black holes with different fluxes.

\subsection{Expansion and physical quantities near the horizon}

The black hole horizon is defined by $f(r_+)=0$. Given a horizon we can proceed to
expand the solution near $r=r_+$. In the coordinates we have chosen, regularity requires the
$g_{tt}$ component of our metric to have a double zero at the horizon, while the $g_{xx}$ component should go to a nonzero constant.  
 Solving the equations of motion (\ref{2nd_order_eqn}) in terms of our Ansatz (\ref{eq:Ansatz}) under these requirements, we find the expansions
\bea\label{bhexpansion}
f(r) & = & f_0\left(\left(1-\frac{r}{r_+}\right)^2+\left(1-\frac{r}{r_+}\right)^3+ \frac{11+16 h_0^2}{12}\left(1-\frac{r}{r_+}\right)^4+\dotsi\right) \,, \\
p(r) & = & p_0\left(1+\frac{5-2 h_0^2}{2}\left(1-\frac{r}{r_+}\right)^2+\frac{5-2 h_0^2}{2}\left(1-\frac{r}{r_+}\right)^3+\dotsi\right)\,,\\
h(r) & = & \sqrt{f_0}\left(h_0\left(1-\frac{r}{r_+}\right)^2+h_0\left(1-\frac{r}{r_+}\right)^3+\dotsi\right)\,.
\eea

Varying the constant $h_0$ produces the family of black holes we will study. Varying $r_+$ only causes various quantities to scale in a way determined by their dimension, and so $r_+$ will drop out of the dimensionless quantites that we will compute. The constants $f_0$ and $p_0$ are unfixed at the horizon and determine the clock and ruler of the system. We will eventually fix them by imposing Lifshitz asymptotics on a scale set by the temperature $T \gg \L^2$ as in (\ref{asymptotics}) and (\ref{asymptotics2}).

We can calculate a few physical quantities characterising the horizon of these black holes, specifically their temperature, entropy density, and horizon flux density. These are given by
\bea
T &=& \frac{r_+}{2\pi}\sqrt{\frac{1}{2} \frac{d^2f}{dr^2}} \Big\lvert_{r=r_+} \,, \\
s &=& 2 \pi \Big(\frac{\ell}{\k}\Big)^2 \, p(r_+) \,, \\
\phi &=& \frac{\ell g r_+}{\k}\left(\frac{p}{\sqrt{f}}\frac{dh}{dr}\right)\Big\lvert_{r=r_+} \,.
\eea
Using the expansion (\ref{bhexpansion}) near the horizon, we obtain
\be
T = \frac{\sqrt{f_0}}{2\pi},\qquad s = 2\pi p_0\left(\frac{\ell}{\k}\right)^2,\qquad \phi = 2 h_0 p_0\left(\frac{\ell g}{\k}\right) \,.
\ee

Additionally, we can compute the value of the $r$-independent quantity $K$ from (\ref{niftycomb}) above, which we can then rewrite in terms of $T$ and $s$:
\be
K=\sqrt{f_0}p_0 = Ts\left(\frac{\k}{\ell}\right)^2 \,.
\ee
From (\ref{1st_law_1}) this immediately implies that
\be\label{firstlaw}
\CF=\CE-Ts,
\ee
which is the integrated form of the first law for these black holes. This result is an important consistency check for our framework and we will use it later as a check on the accuracy of our numerics.

Before turning to numerics, we can make some analytic predictions about these physical quantities in the limit where the marginally relevant mode is not excited.  In this case, when $\Lambda=0$, we expect from Ward identities that the pressure should be equal to the energy, because our anisotropic scale invariance is unbroken.  Recalling that $\CF=-{\cal P}$, we therefore expect that $\CF_0=-\CE_0$.  By examining the expressions for $\CF$ and $\CE$ in (\ref{eq:niceEF}), we see this relation can only be true if $\beta=0$.  Combining with the integrated first law, we thus expect the following results at $\Lambda=0$:
\be\label{zeroresults}
\CE_0=-\CF_0=\frac{1}{2}Ts_0\,,\qquad \CJ^{\hat t}_0=0.
\ee
This relation between $\CE$ and $Ts$ at $\Lambda=0$ is a special case of the relationship for general values of $z$ demonstrated in \cite{Bertoldi2009a}.

\subsection{Integrating towards the Lifshitz Boundary}

Equipped as we now are with various physical expectations and mathematical relations, let us explore the results of numerical integration.  Our basic technique is to take the near horizon expansions in (\ref{bhexpansion}) and then use the equations of motion in (\ref{2nd_order_eqn}) to integrate towards the boundary.  We will always use $r/r_+$ as our variable, and we will want to explore unitless quantities such as $\CF/Ts$, $s/T$ and $\Lambda^2/T$.

When performing this integration, we find that we cannot use a value of $h_0$ bigger than $h_\text{max}\approx .9714$. At larger values of the flux, the numerical integration procedure produces metric functions which grow exponentially as we try to take $r$ to $0$, and we can never reach a boundary. This behaviour was first noted in \cite{Danielsson2008}. While this implies a maximum value of the horizon flux for the black holes we consider, asymptotically we will see that this limiting value corresponds to the high temperature limit $\Lambda^2/T \to 0$, in which the deformation is turned off and we recover a black hole in the pure Lifshitz spacetime. These are the black holes that have been considered in the previous works we mentioned above.

We also have a minimum value of the flux; when the flux is strictly zero at the horizon, we reproduce standard asymptotically AdS black holes in which the flux is zero everywhere. However, the limit in which the flux is taken to be very small is nontrivial. This limit corresponds to the zero temperature limit $\Lambda^2/T \to \infty$. If we take this limit with $\Lambda$ fixed then we recover the RG flow from UV Lifshitz to IR AdS found in \cite{Kachru2008}. In units of the temperature, the crossover regime to Lifshitz scaling moves further and further away from the horizon. This is why if we keep the horizon quantity $T$ fixed in this limit we simply recover pure AdS black holes.

Thus we will find that tuning the horizon flux via $h_0$ corresponds to interpolating between the zero temperature RG flow of \cite{Kachru2008} and the black holes solutions without the marginally relevant deformation of e.g. \cite{Taylor2008, Bertoldi2009, Ross2009, Mann2009, Danielsson2008, Bertoldi2009a}. For the remainder of the paper, for technical reasons to be discussed shortly, we will be most interested in solutions whose flux is just below the maximum; that is, we will be exciting only a small amount of the marginally relevant mode.

\subsubsection{Matching $\Lambda$, $f_0$ and $p_0$}

Before we can attempt to obtain the thermodynamic variables of interest through formulae like (\ref{eq:niceEF}), we need to characterise the asymptotics by extracting $\Lambda$, $f_0$ and $p_0$. The first of these is a physical quantity in its own right, while the latter two will tell us how to rescale space and time in order to consistently compare between different solutions.

The scale $\Lambda$ can be found by fitting the numerical results to the asymptotic expectations.  Concretely, we fit to the expansions in (\ref{series_solutions}) sufficiently close to the boundary that the $r^4$ terms are suppressed. For the solutions we consider, we find it is sufficient to evaluate near $\log r/r_+ \approx -10^5$.  This procedure allows us to find the value of $\Lambda$ for a given solution, provided we fix a value of $\lambda$ in (\ref{series_solutions}). In practice the dimensionless quanitity we obtain is $\Lambda r_+$ which we will be able to relate to $\Lambda^2/T$ once we have found $f_0$.  A quick check we performed on the numerics was to reproduce the `covariance' behavior of (\ref{covariance1}) for $\Lambda$ as a function of $\lambda$. As in previous sections, we will now set $\lambda = 0$.

Upon extracting $\Lambda$ in this manner as a function of $h_0$ we find that $h_0=h_\text{max}$ gives $\Lambda=0$.  Therefore the maximum flux solution is also the solution with none of the marginally relevant mode excited.  We now have a first order understanding of why problems arise when trying to integrate solutions with more flux; we would have to match them onto asymptotics with negative $\Lambda$.

Extracting $f_0$ and $p_0$ is more subtle. These divide out of the equations (\ref{2nd_order_eqn}) and must be fixed by specifying the asymptotic normalisation of the metric components $g_{tt}$ and $g_{xx}$. The leading order logarithmic running to the boundary uncovered in (\ref{asym_metric}) means that the $r \to 0$ limit of $r^4 g_{tt}$ and $r^2 g_{xx}$ remains $r$ dependent. This was manifested in (\ref{eq:rescaling})
where we needed to rescale time and space in a $\mu$-dependent fashion in order to obtain the near-Lifshitz form of the metric (\ref{asymptotics}) and (\ref{asymptotics2}). As we will be working close to $\Lambda = 0$, at small $\Lambda^2/T \ll 1$, we can take $\mu \sim r_+^{-1}$ in  (\ref{asymptotics}) and (\ref{asymptotics2}). In this high temperature regime we thus have a well defined region in which the near-Lifshitz form holds. This will allow us to fix $f_0$ and $p_0$ in the undeformed theory and then extend away from $\Lambda=0$.

\begin{figure}[h]
  \begin{center}
    \includegraphics[width=2.7in]{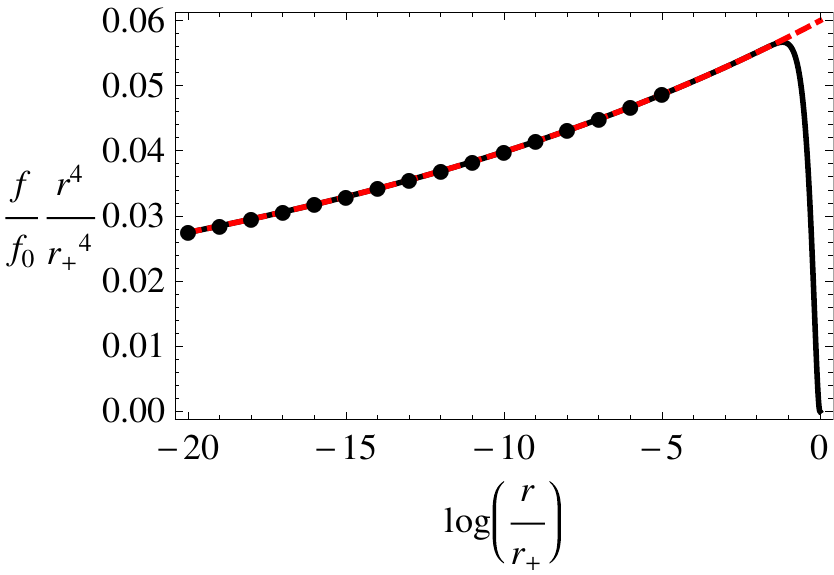} \hspace{0.3cm}
    \includegraphics[width=2.7in]{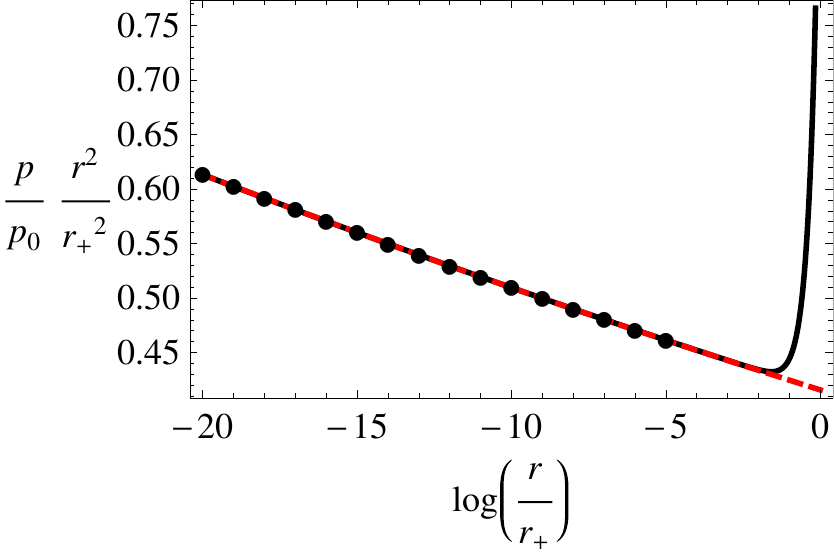}
  \end{center}
  \caption{\small Left: $f r^4/f_0 r_+^4$ versus $\log r/r_+$. The dashed red line shows the function $f r_+^4/f_0 r^4$ as given in (\ref{asymptotics}), including higher order terms, for $f_0 r_+^4=27.72$. Right: $p r^2/p_0 r_+^2$ versus $\log r/r_+$, with the dotted line corresponding to $p_0 r_+^2=1.89$. The black line is the numerical curve and the black dots are the values used for fitting.}
  \label{f0p0vsrho}
\end{figure}

In figure \ref{f0p0vsrho} we illustrate how the numerically generated $f$ and $p$ can be fitted over a finite range, at high temperatures, to the near-Lifshitz form (\ref{asymptotics}) and (\ref{asymptotics2}). In fact we used these expansions to fifth order in $1/\log (\Lambda r_+)$. This allows us to obtain $f_0$ and $p_0$. The figure depicts the fitting procedure for $h_0=.962$, which corresponds to $\log \Lambda^2/T\approx -200$. The fitting is even more robust for the smaller values of $\Lambda$ which we will usually consider.

\begin{figure}[h]
  \begin{center}
    \includegraphics[width=3in]{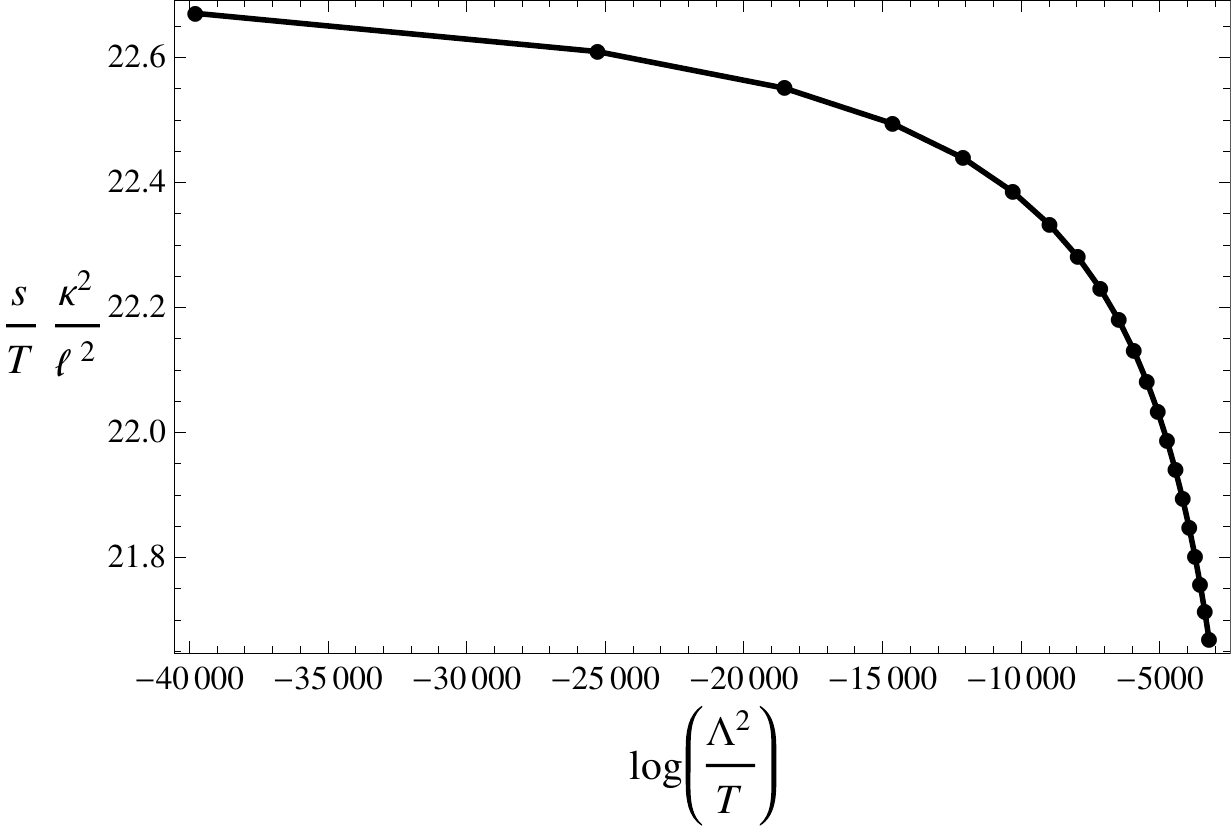}
  \end{center}
  \caption{\small Entropy density over $T$ versus $\log \Lambda^2/T$ for $h_0=0.9713$ to $h_0=0.9708$. Dots are data points, which are joined by straight lines.}
  \label{soverT}
\end{figure}

Having found $\Lambda$, $f_0$ and $p_0$ for a solution with a given $h_0$, we can plot the (dimensionless) entropy density over temperature as a function of $\log(\Lambda^2/T)$. This is shown in figure \ref{soverT}. In particular at $\Lambda=0$ we find
\be
\frac{s}{T} \Big\lvert_{\Lambda=0}\approx \frac{4 \pi^2}{\sqrt{3}}\left(\frac{\ell}{\k}\right)^2 \,.
\ee
We have only found this value numerically, to several digits of precision, and do not have an {\it a priori} understanding of its origin, since we do not know the form of the analytic solution. 
The value we find 
seems to agree with the value $s/T \approx 11.4$ quoted in \cite{Danielsson2008} if we set $(\ell/\kappa)^2 = \half$.


\subsubsection{Energy, Free Energy, and $\CJ^{\hat{t}}$}

Our next objective is to compute the temperature dependence of $\CF/T s$, $\CE/T s$ and $\CJ^{\hat{t}}/T s$ with the numerators given in (\ref{eq:Ftotal}), (\ref{energy}), and (\ref{gauge}) respectively. As well as being unitless, these ratios are independent of the constants $f_0$ and $p_0$ and are therefore not sensitive to any, possibly delicate in this logarithmically running theory, choice of normalisation of time and space.

\begin{figure}[bt]
  \begin{center}
    \includegraphics[width=1.8in]{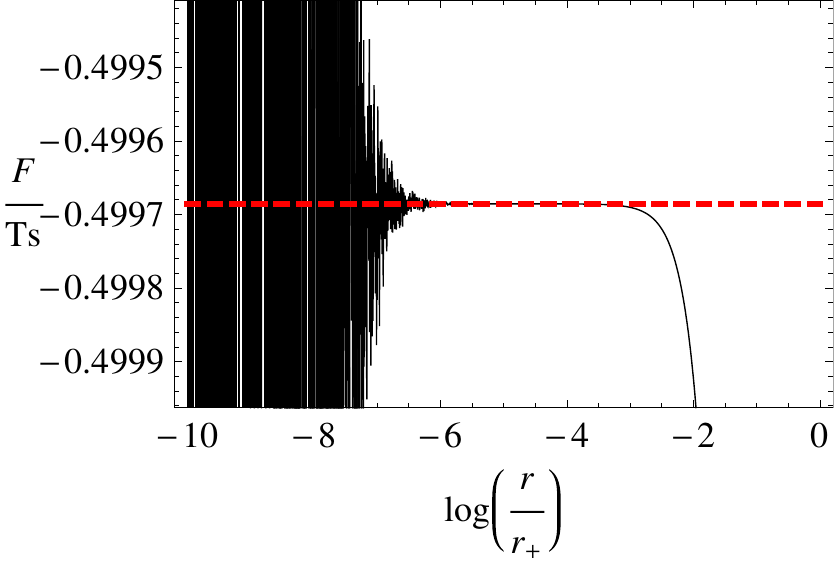}\hspace{0.2cm}
    \includegraphics[width=1.8in]{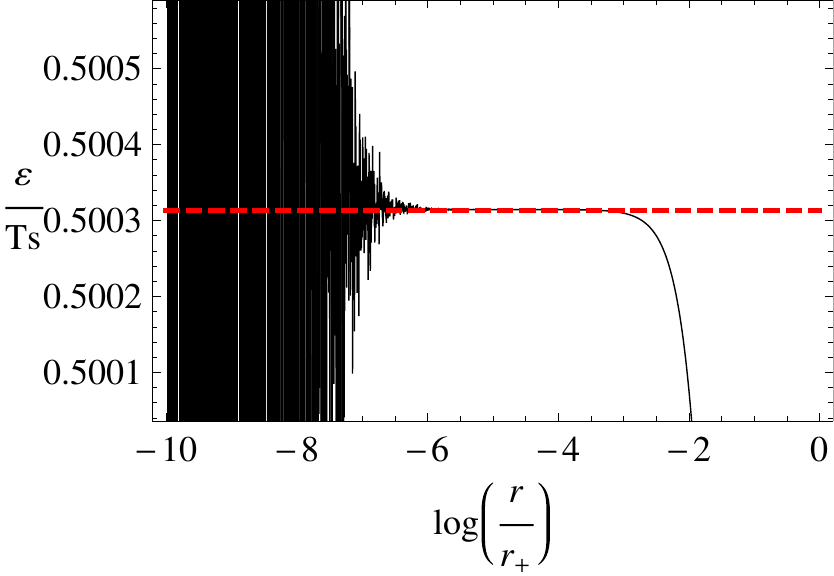}\hspace{0.2cm}
    \includegraphics[width=1.8in]{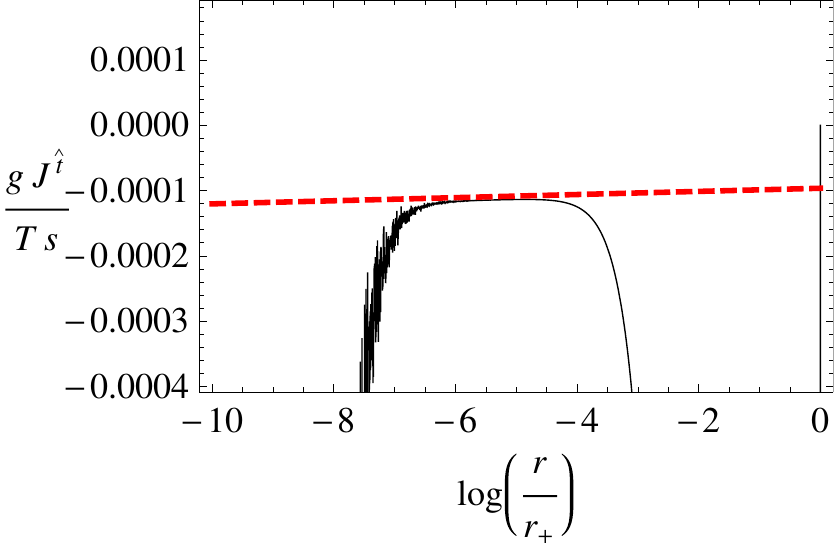}
  \end{center}
  \caption{\small Plots of $\CF/Ts$, $\CE/Ts$ and $\CJ^{\hat t}/Ts$ as a function of $\log r/r_+$ for $h_0=.971$, corresponding to $\log (\Lambda^2/T) \approx -3210$. The near horizon region is dominated by IR effects while near the boundary divergences set in due to use of a truncated series of counterterms (\ref{eq:coeffs}). The quantities are well defined in the intermediate region, which can be made large by working with a sufficiently high order counterterm expansion. Black lines are numerical results while the red dashed line is a fit to the analytical renormalised expressions.}
  \label{fejvsrho}
\end{figure}

We cannot directly extract $\a$ and $\beta$ from the numerics and then use the expressions (\ref{eq:niceEF}) for the energy etc., because the $r^4$ terms in the near boundary expansions (\ref{series_solutions}) are exponentially small compared to the logarithmically decaying modes. Instead we will use the expressions (\ref{eq:Ftotal}), (\ref{energy}), and (\ref{gauge}) together with the series expansions for the counterterm coefficients in (\ref{eq:coeffs}). In practice we only know the counterterm expansion (\ref{eq:coeffs}) up to some fixed (arbitrarily high in principle) order. The main fact that restricts us to the high temperature regime is that the truncated counterterm series is more accurate there. With $\Lambda^2/T \ll 1$ we can estimate that truncating the expansions (\ref{eq:coeffs}) at order $1/\log^N (\Lambda r)$ will be reliable for
\be
\left| \log \frac{r}{r_+} \right| \lesssim \frac{N}{4} \log \left(-\log\frac{\Lambda^2}{T} \right) \,.
\ee
We see that increasing the number $N$ of terms allows us to move closer the boundary, as does increasing the temperature. At sufficiently large $N$ and temperature we can expect to obtain a constant intermediate region between the near horizon IR effects and the asymptotic divergences. From this region we can read off our thermodynamic variables. This process is illustrated in the plots of figure \ref{fejvsrho}, for which we expanded the counterterms to ten orders.

There is an additional complication in obtaining $\CJ^{\hat{t}}$. For $\CF$ and $\CE$ the counterterms not only remove all divergences, but also all subleading terms in $1/\log(r/r_+)$. Therefore the only significant error is due to truncating the counterterm expansion. This is not true for $\CJ^{\hat{t}}$; while the counterterms remove the divergences, there are still inverse powers of $\log r/r_+$ that vanish asymptotically but which are nontrivial in the regime in which we can read off the value of $\CJ^{\hat{t}}$. The behaviour can be derived analytically from (\ref{gauge}) and so we can read off $\CJ^{\hat{t}}$ accurately by fitting the numerics to an expansion of several orders in $1/\log(r/r_+)$ in the intermediate regime. Finally, we should recall that the quantity $\CJ^{\hat{t}}$ is dependent upon ambiguities in the counterterms and is not a conserved charge.



\subsection{Exploring the dependence of $\CE$ and $\CF$ on $\log \Lambda^2/T$}

\begin{figure}[h]
  \begin{center}
    \includegraphics[width=1.8in]{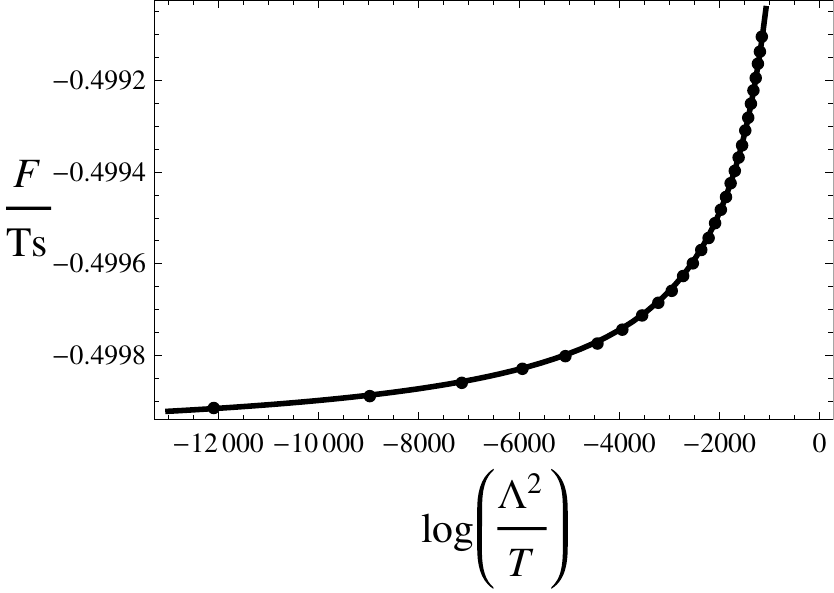}\hspace{0.2cm}
    \includegraphics[width=1.8in]{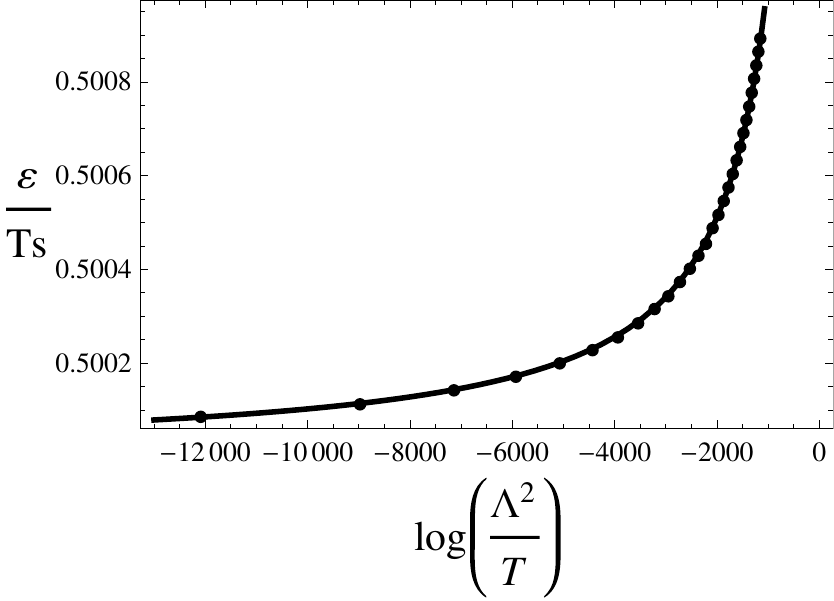}\hspace{0.2cm}
    \includegraphics[width=1.8in]{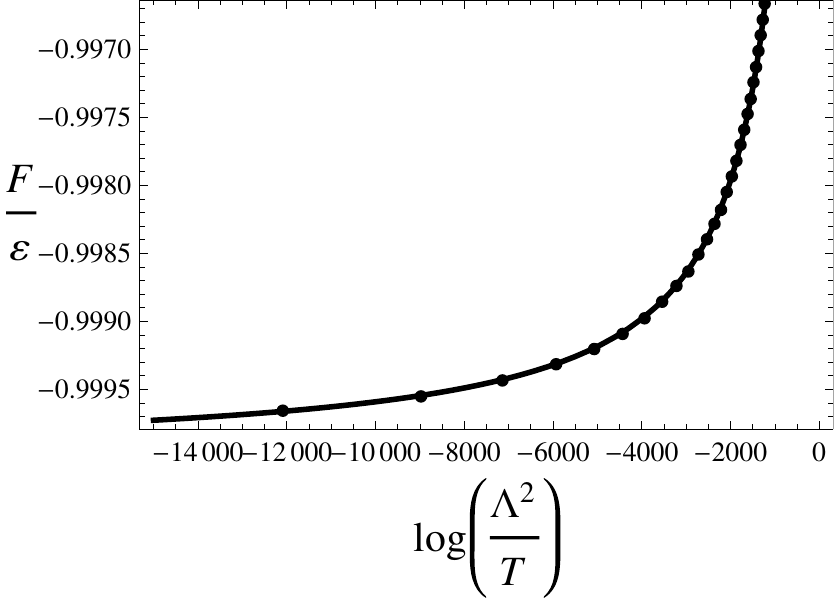}
  \end{center}
  \caption{\small Plots of $\CF/Ts$ and $\CE/Ts$ , and $\CF/\CE$, as functions of $\log (\Lambda/T^2)$.
  The range of $h_0$ is from $0.9712$ to $0.9698$, corresponding to $\log \Lambda^2/T$ from about $-12000$ to $-1150$.
  The dots are numerical results, while the lines are the fits in equations (\ref{FoverSTfit}), (\ref{EoverSTfit}), and (\ref{FoverEfit}).}
  \label{FandEoverST}
\end{figure}

We finally have all of the tools available to compute $\CF/Ts$ and $\CE/Ts$ as a function of $\log \Lambda^2/T$. The results are shown in figure 
\ref{FandEoverST} for a range of high temperatures.

The first observation to make about figure \ref{FandEoverST} is that we reproduce the anticipated results (\ref{zeroresults}) as
$\Lambda \to 0$ (that is, as $\log \Lambda^2/T \to - \infty$).  We find numerically
\be
\frac{\CF_0}{s_0T}=-\frac{1}{2},\qquad \frac{\CE_0}{s_0T}=\frac{1}{2},\qquad \frac{\CF_0}{\CE_0}=-1\, .
\ee
Beyond this limit, we find stable fits for each of these quantities as an expansion in inverse powers of $\log\Lambda^2/T$.  For $\CF$ we find
\be\label{FoverSTfit}
\frac{\CF}{Ts}=-\frac{1}{2}+\frac{1}{\log \Lambda^2/T}+\dotsi \,.
\ee
The coefficent of $1$ in the numerator of the second term here is not known analytically, it is obtained to some precision from our numerics. The fit (\ref{FoverSTfit}) is also shown in figure \ref{FandEoverST}. The simplicity of this term may suggest the existence of a simple analytic solution.

We can now use the integrated first law (\ref{firstlaw}) to predict the expansion for $\CE$:
\be\label{EoverSTfit}
\frac{\CE}{Ts}=\frac{1}{2}+\frac{1}{\log \Lambda^2/T}+\dotsi \,,
\ee
which, as Figure (\ref{FandEoverST}) shows, is strongly supported by the numerical results.  Note that $\frac{\CE}{Ts}$ is completely determined by $\frac{\CF}{Ts}$, even beyond the leading order terms written down in (\ref{FoverSTfit},\ref{EoverSTfit}).
Combining the expansions (\ref{FoverSTfit}) and (\ref{EoverSTfit}), we expect $\CF/\CE$ to fit
\be\label{FoverEfit}
\frac{\CF}{\CE}=-1+\frac{4}{\log \Lambda^2/T}+\dotsi \,,
\ee
which is also seen to hold within numerical precision for this range of $\log \Lambda^2/T$.

\begin{figure}[h]
  \begin{center}
    \includegraphics[width=3.3in]{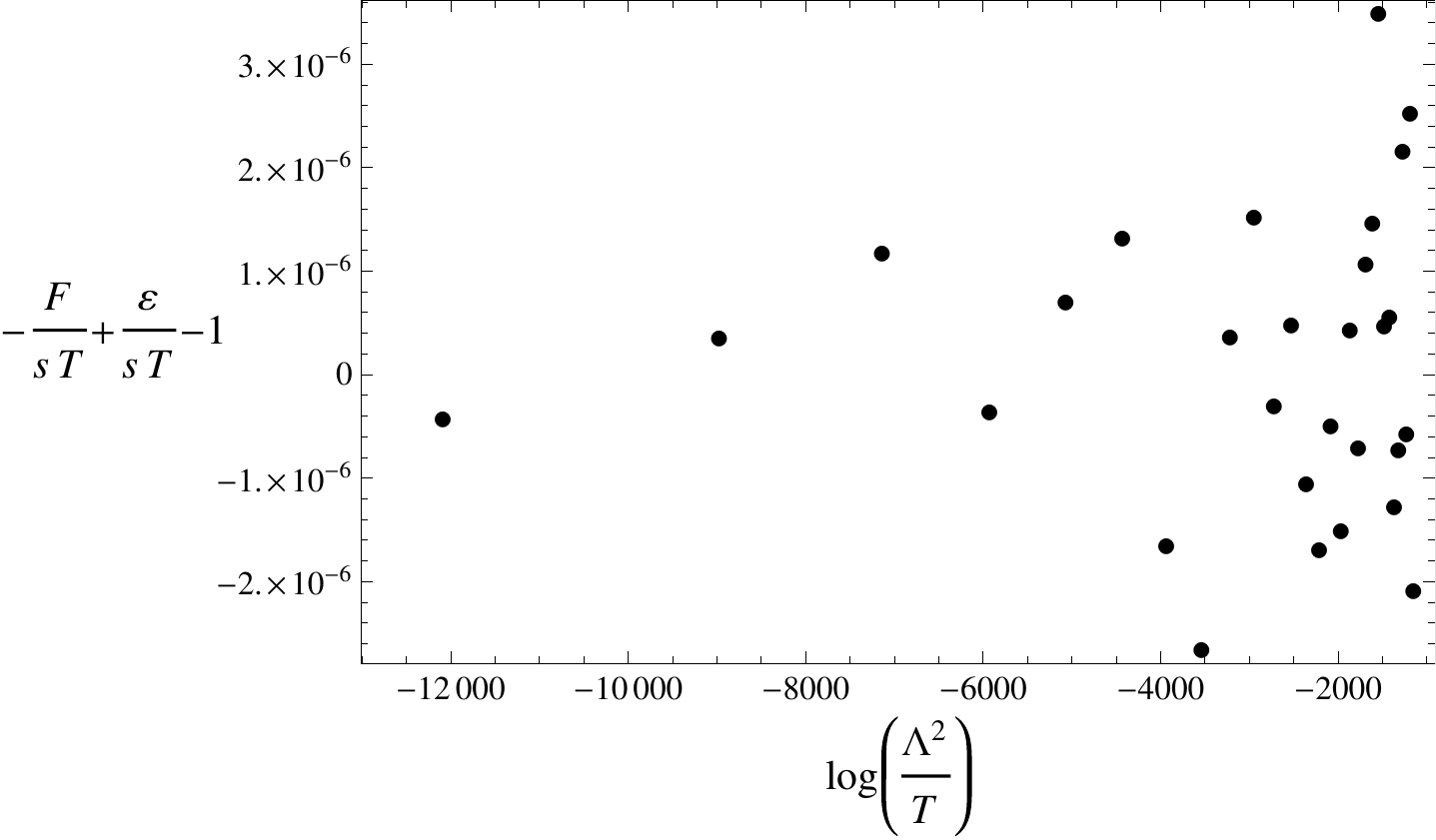}
  \end{center}
  \caption{\small Scatter plot of $-\CF/Ts+\CE/Ts-1$ for $h_0$ from $0.9712$ to $0.9698$, showing that the integrated first law is obeyed numerically to good accuracy.}
  \label{firstlawcheck}
\end{figure}

Thus the graphs in figure \ref{FandEoverST} provide a check for the integrated first law, or rather that our numerics satisfy the law. In order to see how precisely our numerics satisfy this requirement, we can examine figure \ref{firstlawcheck}, which plots $-\CF/Ts+\CE/Ts-1$.  Since we expect this quantity to be zero, its size gives us an estimate of our numerical error.  As the plot shows, our error is less than $1$ part in $10^5$ for the range considered. While the variation of our variables of interest in figure \ref{FandEoverST} are also small, the variation within the range of $\L^2/T$ we consider is at least two orders of magnitude bigger than the error. Furthermore the error does not appear to be systematic.

\begin{figure}[bt]
  \begin{center}
    \includegraphics[width=3.3in]{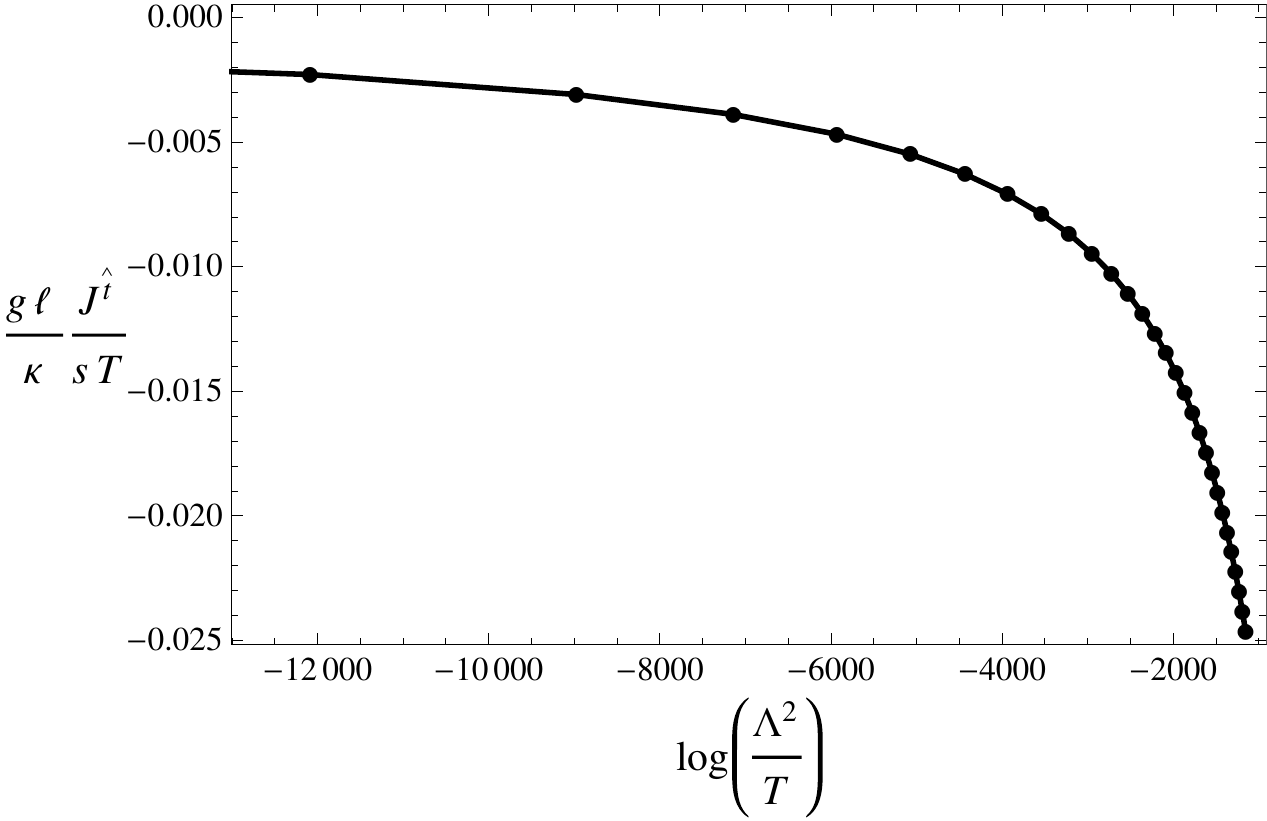}
  \end{center}

  \caption{\small $\CJ^{\hat{t}}/Ts$ for $h_0$ from $0.9712$ to $0.9698$. Dots are numerical data points joined by lines.}
  \label{JhatoverST}
\end{figure}

Lastly, we present a plot of $\CJ^{\hat{t}}/Ts$ against $\log \Lambda^2/T$, in figure \ref{JhatoverST}. As we can see from the figure, we find that
$\CJ^{\hat{t}}=0$ at $\Lambda \to 0$ as expected. We have not been able to fit this curve to a simple function.

\pagebreak

\section{Discussion}

Our main concrete result is perhaps equation (\ref{FoverSTfit}) expressing the $\Lambda^2/T$ dependence of the free energy of the theory upon approach to Lifshitz quantum criticality at $\Lambda = 0$. It is important to characterise the strongly coupled physics of (marginally) relevant deformations away from criticality. To the extent that such deformations are generic, they should be included in attempts at holographic model building for condensed matter systems along the lines of \cite{Hartnoll:2009c}. Beyond the thermodynamics we have studied in this paper, it will be of interest to consider correlators of fields in the deformed background and describe their behaviour upon approach to criticality.

More generally, however, we have tried to grapple with the technical challenge of renormalising the leading order logarithmic running of the deformation of the theory generating the scale $\Lambda$. We have worked perturbatively about the $\Lambda=0$ point (equivalently, at high temperatures relative to $\Lambda$) and successfully computed finite thermodynamic quantities. Obtaining these results required a certain amount of numerical acrobatics and it would clearly be advantageous to have a better formal understanding of the spacetime asymptopia as well as a more general approach that would allow computations all the way to the complementary $T=0$ limit.

While the deformation dynamically generating the scale $\Lambda$ has many of the hallmarks of a marginally relevant deformation of a Lifshitz invariant theory (e.g. at $T=0$ the theory runs to a new IR fixed point while for $\Lambda^2 \ll T$ one recovers scale invariant results), we have not shown rigorously that our spacetimes can be considered `asymptotically Lifshitz'. In fact, the leading order logarithms in Fefferman-Graham-like coordinates (\ref{asym_metric}) may indicate that this is not strictly the case. The need to rescale space and time together with the energy scale in (\ref{eq:rescaling}) may suggest a weak generalisation of the notion of a fixed point\footnote{Thanks to Matt Headrick for discussions of this possibility.}. We hope that future work will elucidate this question.

In weakly coupled realisations of $z=2$ Lifshitz symmetry, reviewed briefly at the end of section 3 above, the relevant operator that drives the theory to a relativistic IR fixed point is also responsible for the existence of `tilted'  phases breaking spatial isotropy. A fascinating open question is to identify such phases, if they exist, in the gravitational dual. One is led to wonder if such phases exist beyond the maximal horizon flux that we found.

There are various further natural extensions of our work: An analysis similar to ours should be possible, perhaps easier, for $z>2$ where the deformation mode becomes strictly relevant. It would be very helpful to have analytic solutions, we found some encouraging signs that this might be possible. Finally, logarithmic modes can occasionally make the stability of the spacetime quite a subtle question -- it would be interesting to characterise the fluctuations about our marginally deformed backgrounds.

\section*{Acknowledgements}

We would especially like to thank Matt Headrick for several clarifying long discussions. We have also benefited from stimulating discussions with
Dionysios Anninos, Vijay Balasubramanian, Sophie de Buyl, Stephane Detournay, John McGreevy, Mike Mulligan, Balt van Rees, Omid Saremi,  Andy Strominger, Wei Song, Tom Hartman and Marika Taylor. This work of MCNC was supported in part by DOE grant DE-FG02-91ER40654 and the Netherlands Science Organisation (NWO). The work of SAH is partially supported by DOE grant DE-FG02-91ER40654 and by the FQXi foundation. CAK is supported by the Fundamental Laws Initiative at the Center for the Fundamental Laws of Nature.

\bibliography{ref_list_lif_dec12}{}

\end{document}